\documentclass[jgrga]{agu2001}        %preprint version
\usepackage{epsfig}

\authorrunninghead{TURCOTTE \& WIMMER-SCHWEINGRUBER}

\titlerunninghead{IN SITU TESTS FOR SOLAR CHEMICAL EVOLUTION}

\journalid{}
\articleid{}{}
\paperid{}
\cpright{}{}
\received{} \revised{} \accepted{} \published{}

\authoraddr{S. Turcotte, Lawrence Livermore National Laboratory, L-413,
P.O. Box 808, Livermore, CA 94551, USA. (sturcotte@igpp.ucllnl.org)}
\authoraddr{R. F. Wimmer-Schweingruber, Physikalishes Institut,
Universit\"{a}t Bern, Sidlerstrasse 5, CH-3012 Bern, Switzerland.
(robert.wimmer@phim.unibe.ch)}

\begin{document}

%%%%%%%%%%%%%%%%%%%%%%%%%%%%%%%%%%%%%%%%%%%%%%%%%%%%%%%%%%%%%%%%%%%%%%%%%%
%
%Other possibility for title:
%\title{Elemental Evolution in the Sun and Possible in situ Tests}
\title{Possible in situ Tests of the Evolution of Elemental and Isotopic
Abundances in the Solar Convection Zone}

\author{S. Turcotte}
\affil{Lawrence Livermore National Laboratory, Livermore, CA, USA}

\author{R. F. Wimmer-Schweingruber}
\affil{Physikalishes Institut, Universit\"{a}t Bern, Bern, Switzerland}

\begin{abstract}
Helioseismology has shown that the chemical composition of the Sun
has changed over its lifetime. The surface abundance of helium and heavy
elements is believed to have decreased by up to 10\% relative to their initial
values. However, this reduction is too small to be tested by direct observations of
the photospheric chemical composition. Here, we compare the predicted variations in
the solar photospheric composition with precise measurements of abundances 
in meteorites and the solar wind composition.
Although elemental composition ratios can vary by roughly a percent (e.\,g.\,for Ca/Mg and Ca/Fe)
over the Sun's lifetime,
their measurements are rife with uncertainties related to uncertainties in
the interpretation of meteoritic measurements, photospheric determinations, and the complex fractionation
processes occurring between the upper photosphere and lower chromosphere and the corona. 
On the other hand, isotopic ratios can be measured much more
accurately and are not expected to be affected as much by extrasolar processes,
although more work is required to quantify their effect.
As the isotopic ratios evolve in the Sun proportionally to the mass ratios
of the isotopes, light elements yield the highest variations in isotopic ratios.
They are predicted to reach as high as 0.6\% for $^{18}$O/$^{16}$O and are only slightly
lower in the cases of $^{26}$Mg/$^{24}$Mg and $^{30}$Si/$^{28}$Si. Such a value
should be well within the sensitivity of new missions such as Genesis. 
\end{abstract}

\begin{article}
\section{Introduction}
 
Modern solar models can be tested to a high degree of accuracy with
newly available helioseismic data.
Such models infer the speed of sound in the solar interior by comparing
measured
oscillation modes with model calculations. Because the speed of sound
depends on the mean molecular weight
of the gas (among other quantities), determining profiles of sound speed
yields information about possible
compositional inhomogeneities in the interior of the Sun. Indeed, solar
models which include effects of
elemental segregation succeed much better than those without in
reproducing inferred speed profiles
[{\it Christensen-Dalsgaard et al.}, 1996]. In these models, 
the photospheric abundances
of all elements heavier than hydrogen
have decreased by up to about 10\% over the history of the Sun. However,
this settling of the heavier elements out of the outer convection zone
toward the center of the Sun has not
been detectable with photospheric abundance measurements. In this work
we consider the possibility of directly
measuring this effect by measuring the composition of the solar wind.
 
Lunar soils have preserved an archive of solar wind data dating back to
the time of the formation of the lunar regolith.
Dust grains that have been exposed to the solar wind retain it with
variable fidelity, although the lunar
regolith as a whole appears to have preserved the entire solar wind
fluence incident on the lunar surface
[e.\,g.\,{\it Wieler}, 1998]. Lunar soils are especially well suited to
measure the isotopic and, to a limited extent, the
elemental composition of the noble gases and of few other elements,
notably nitrogen. Early measurements
of the He isotopic composition appeared to show a secular variation of
the solar wind $^3$He/$^4$He ratio [{\it Geiss}, 1973].
This could be interpreted as due to differences in the solar wind
acceleration in the past or due to secular admixture of
$^3$He from the $^3$He bulge in the Sun by a slow mixing process beneath
the outer convection zone [{\it Bochsler et al.}, 1990].
The apparent secular change of the isotope ratio $^3$He/$^4$He has
recently been re-interpreted.
{\it Heber} [2002] analyzed lunar soil material of various antiquities for
their noble gas isotopic composition.
Apart from the known changes in He isotope composition, they found a
change in the Ne isotope composition which is considerably larger than
what could be attributed to the chemical evolution of photospheric
composition.
According to that work, the apparent secular change is due to slow
removal of surface layers of lunar soil grains. The older
grains have had more surface material removed, increasing the importance
of the deeper-seated and isotopically heavier
so-called ``SEP'' or ``HEP'' component
[{\it Wieler et al.}, 1986; {\it Wimmer-Schweingruber and Bochsler}, 2001a]. 
{\it Heber} [2002] used the measured correlation
between Ne and He isotope composition to account for this removal by
assuming that the Ne isotope composition
did not change over time. Then the apparent secular variation of the He
isotopes vanishes and hence the variations of $^3$He/$^4$He are
not necessarily a test for solar evolution models. Nitrogen is known to
have large
secular variations in its isotopic composition in lunar soils [see
e.\,g.\,{\it Kerridge}, 1993, for a review.].
However, recent investigations by {\it Wieler et al.}~[1999] have shown that
about 90\% of the implanted nitrogen is non solar,
i.\,e.\,due to some other, as yet unidentified, source.
 
\begin{figure}
     \epsfig{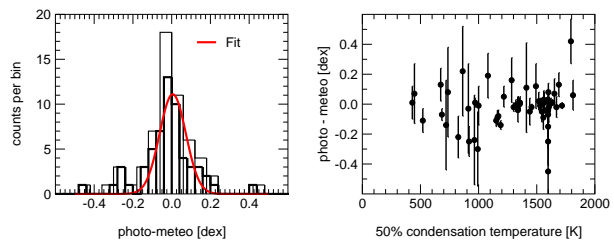}
      \caption{Comparison of CI - meteoritic and
               photospheric abundances normalized to Mg. The right-hand panel is a plot
               of the logarithm of the CI - meteoritic abundance ratio X/Mg divided by
               its corresponding photospheric value plotted versus 50\% condensation
               temperature. The left-hand panel is a histogram of the abundance values
               for two different bin sizes. also shown are a fit to the histogram and the
               uncertainty of the currently best solar wind abundance determination of
               N/O. The uncertainties (and hence the width of the histogram) are nearly
               entirely due to uncertainties in photospheric abundance measurements.
               For instance, atomic parameters limit measurements to an uncertainty of
               about 20 - 30 \%. Meteoritic uncertainties are smaller, on the order
               of 3 - 5\% and mainly due to possible sampling biases.
               \label{fig:met_photo_hist}}
   \end{figure}

Because there is no unambiguous evidence in lunar soils for secular
changes in solar wind composition which could be
interpreted as due to elemental segregation, another approach needs to
be chosen. Here we investigate the possibility
of comparing the composition of certain, well determined meteoritic
abundance ratios with the same abundance ratios
measured in the solar wind.
The formation of the solar system about 4.6 billion years ago left
compositional signatures in
its constituent bodies and in different samples of these bodies. 
Compositional differences can be used as
tracers for formation and alteration processes active mainly in the
early solar system, but partially
continuing to the present day. For example, volatile elements are
extremely depleted in terrestrial samples
and in most meteorites. Elements differ in their chemical properties,
resulting in variations
of elemental abundance ratios in different separates of various
solar-system samples. 
Because the Sun contains 99.9 \% of the mass in the solar system, its
composition serves as the reference for solar
system composition and often solar and solar-system composition are used
synonymously.
This information comes from spectroscopic abundance determinations of the solar
photosphere or from in-situ measurements of the solar wind. However,
much of the information available today comes from studies of the
composition of meteorites and of meteorite mineral separates or samples
from other solar-system bodies. Based on the comparison of
such investigations the composition of the bulk solar system has become
known to a remarkable degree of accuracy. In fact, because of
the close match in elemental abundances between solar and CI-meteoritic
values, CI meteorites are considered the most primitive
material in the solar system besides comets or the Sun. Figure~\ref{fig:met_photo_hist}
illustrates this good agreement using data from the compilation
of {\it Grevesse and Sauval} [1998]. The right-hand panel is a plot of
the logarithm of the CI-meteoritic abundance ratios
of element X with respect to Mg, X/Mg, divided by its corresponding
photospheric value plotted versus 50\% condensation temperature [{\it Wasson},
1985]. This quantity is often used to order elemental abundance trends
in different classes of meteorites. The comparison of
photospheric and CI-meteoritic abundance ratios is not reliable for
condensation temperatures below about 400~K. For all other elements,
with the exception of Li, the agreement between the two different
abundance measurements (CI-meteoritic and photospheric)
is considered remarkably good. This is
illustrated in the left-hand panel of Figure~\ref{fig:met_photo_hist}. It shows a
histogram of the abundance ratios, (X/Mg)/(X/Mg)$_{\rm photo}$. The
histogram is symmetric around zero, indicating that there is no systematic
disagreement between the two sources of solar abundances. The standard
deviation of the histogram is 0.084 dex, corresponding to a width
of 21.5\%. A Gaussian fit with that value for $\sigma$ is plotted for
comparison. The fit is satisfactory, and the result does not change
when using a different bin size for the histogram, e.\,g.\,the one shown
in thin lines. Assuming the width of the histogram to be
representative of the quadratic sum of the uncertainties of
CI-meteoritic and photospheric we subtract quadratically the maximum uncertainty
of CI-meteoritic measurements, 10\%, to obtain the expected uncertainty
of photospheric abundance determinations, $\sim 20\%$. This value
is less than that given by {\it Del Zanna et al.} (2001), 25\% - 30\%, who
consider this to be the uncertainties in atomic properties alone. Moreover,
uncertainties in the treatment of the thermodynamics of the
line-emitting regions in the photosphere and lower chromosphere can be substantial
[{\it Holweger}, 2001]. Nevertheless, some authors do give smaller
uncertainties for certain elements with well known atomic parameters.
Meteoritic uncertainties are generally smaller, on the order of 3\% -
10\% and are mainly due to possible sampling biases and will be
considered later.
Thus, from this discussion, we conclude that abundance ratios of
refractory, i.\,e.\,non-volatile, elements
offer the best choice of measuring the effect of elemental segregation.

%----------------------------------------------------------
% {\bf Caption for Fig~1} {\em Comparison of CI - meteoritic and
% photospheric abundances normalized to Mg. The right-hand panel is a plot
% of the logarithm of the CI - meteoritic abundance ratio X/Mg divided by
% its corresponding photospheric value plotted versus 50\% condensation
% temperature. The left-hand panel is a histogram of the abundance values
% for two different bin sizes. also shown are a fit to the histogram and the
% uncertainty of the currently best solar wind abundance determination of
% N/O. The uncertainties (and hence the width of the histogram) are nearly
% entirely due to uncertainties in photospheric abundance measurements.
% For instance, atomic parameters limit measurements to an uncertainty of
% about 20 - 30 \%. Meteoritic uncertainties are smaller, on the order
% of 3 - 5\% and mainly due to possible sampling biases.}
%----------------------------------------------------------

We consider the physics of elemental segregation in section
\ref{sec:segre}, apply this to solar
evolution in section \ref{sec:evol}, where we also give predictions for
the evolution of some elemental
and isotopic abundance ratios. We consider possible fractionation
processes which would influence
measurements in the solar wind and investigate
the potential of various solar wind
measurements in section \ref{sec:frac}.

\section{The Physics of Chemical Evolution in the Sun}\label{sec:segre}
 
The basic structure of the Sun consists of a dynamically stable
inner sphere which extends to a little more than 70\% of the Sun's
radius
and an overlaying convective shell which extends to
the photosphere.
The turbulent motions in the convection zone are so rapid compared to
other time scales
that one can assume that it is well mixed,
leading to a homogeneous composition in its entirety.
Compositional changes due to nuclear processing in the convection zone
can be excluded
because the temperature is too low. Hence changes in the chemical
composition
have to be generated at the boundaries of the convection zone, either
in the dynamically stable
region at its base or at the solar surface (through infall of material
or mass loss).
As any excess or deficit of an element is diluted in the
entire convection zone, a variation of, say, 10\% in the
observed abundance entails that the mass of the element $i$ in question
added or subtracted from the convection zone must be 
$M_i = 0.1\, X_i\, M_{CZ}$, $X_i$ being the original mass fraction of 
element $i$ and $M_{CZ}$ being the mass of the convection zone (in the current Sun
$M_{CZ}$ is roughly 2\% of the total mass of the Sun [{\it
Christensen-Dalsgaard et al.}, 1996]). For instance, if one terrestrial
planet had migrated into the Sun in the early stages of the formation of the 
solar system, this would have resulted in an increase of solar metallicity 
by only about 1\% using the present-day mass fraction of the convective zone. 
Note that this process will not alter the relative abundances of refractive 
elements among themselves.
Moreover, in the early Sun, the convection zone was deeper and hence more massive 
and a larger flux of matter into it would have been necessary for any observable
effect to be expected.
 
The current mass loss or accretion rates are much too low to have any
measurable impact on photospheric abundances.
Earlier in the Sun's life it is probable that
mass loss or accretion rates were higher.
Because of the faster rotation of the Sun in the past, dynamo action was
enhanced. This made more energy in the form of magnetic fields available
to drive the solar wind. Simple calculations show that this could at
most have doubled
the solar wind fluence over the lifetime of the Moon [{\it Wimmer-Schweingruber and Bochsler}, 2001b], commensurate with
measurements
of solar wind noble gases in lunar soils [{\it Geiss}, 1973].
Extrapolating these results
to earlier times does not result in a significant increase of mass loss.
Moreover, the convection
zone was also much more massive in the past. During the Sun's
evolution on the pre main sequence when mass exchange would be expected
to be greatest the Sun is thought to have been completely convective [{\it Hayashi}, 1961].
The hypothesis of large mass loss rates in the young Sun has been
examined but is not favored by helioseismology [{\it Morel et
  al.}, 1997]. The addition of material that has been
processed by
nuclear reactions in the dynamically stable core can also be excluded.
Again, temperatures are high enough only in its innermost regions making any
exchange at its boundary impossible (with the sole exception of lithium, see
section~\ref{sec:mixing}).
 
\subsection{Migration and Separation Processes}
 
The evolution of abundances in the solar convection zone and in the outer 
radiative envelope just below it
occurs chiefly as a result of the differential effects of gravity on
the elements of different mass. In a simple two component gas, constituted
of hydrogen and helium, the inward gravitational pull
is four times greater on helium than on hydrogen. As a result,
helium will gradually settle toward the Sun's center and
the fractional abundance of hydrogen will rise at the surface.
Differential radiation pressure induces a minor correction in the overall
chemical evolution in the convection zone but is plays a dominant role
in the evolution of abundances ratios. 
 
The rate at which gravitational settling occurs is  determined
in a large part by the ``friction'' within the gas, which is dominated by Coulomb interactions.
Because friction increases with the overall ionization state
of the plasma, the settling rate generally will decrease as temperature
and density rise.
 
As the evolution of photospheric abundances is governed by the
evolution of the abundances at the base of the convection zone, the
deeper the convection zone, or the smaller the stable radiative interior, 
the smaller are the fluxes of elements in and out of the convection zone. 
This is the combined effect of enhanced interactions between particles leading to smaller
diffusion velocities and the geometrical effect of reducing the area of the
surface through which particles can flow.
Moreover, the dilution of abundance changes
in the larger convection zone
further increases the time scale for the rate of change of composition.
This is important because the depth of
convection zone varied strongly as the Sun evolved [as shown in {\it Bahcall
et al.}, 2001].
 
\begin{figure}
     \epsfig{file=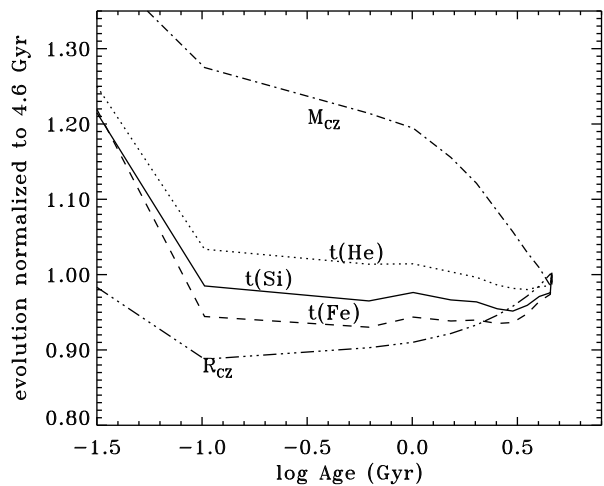,width=8cm}
      \caption{Evolution of the timescale for the variation of
                 chemical composition in the convection zone with the
  age of the Sun, the
                 values are normalized to those in the contemporary Sun
                 at $\log$(age)$=$0.66.
                 The timescale is element dependent, here helium (t(He),
  normalized to 39 Gyr)),
                 silicon (t(Si), normalized to 51 Gyr) and iron (t(Fe),
  normalized to 51 Gyr) are shown.
                 Also shown are the evolution of the mass of the
  convection zone ($M_{CZ}$, normalized at the current value of
  0.023~M$_\odot$), and the radius of the base of the convection zone
  ($R_{CZ}$, which is currently at 0.710~R$_\odot$ in this model).
              }
         \label{fig:timescale}
   \end{figure}

The time scale for the evolution of surface abundances can be
approximated as [{\it Michaud et al.}, 1976] 
\begin{equation}
  \tau = {M_{CZ}\over 4\pi\,\left(r^2\rho v_{\rm diff}\right)_{CZ}}\ \,
,
\end{equation}
where $\rho$ is the density, $r$ is the radius, and $v_{\rm diff}$ is the diffusion
velocity all evaluated 
at the base of the convection zone (CZ).
The timescales for helium, silicon and iron with respect to time 
shown in Figure~\ref{fig:timescale} illustrate the combined effect of the
increases in $v_{\rm diff}$ and the radius of the base of the convection zone 
and the reduction in $M_{CZ}$.
 
%----------------------------------------------------------- surface
% {\bf Caption for Fig~2} {\em Evolution of the timescale for the variation of
% chemical composition in the convection zone with the age of the Sun, the
% values are normalized to those in the contemporary Sun at $\log$(age)$=$0.66.
% The timescale is element dependent, here helium (t(He), normalized to 39 Gyr)),
% silicon (t(Si), normalized to 51 Gyr) and iron (t(Fe),
% normalized to 51 Gyr) are shown.
% Also shown are the evolution of the mass of the
% convection zone ($M_{CZ}$, normalized at the current value of
% 0.023~M$_\odot$), and the radius of the base of the convection zone
% ($R_{CZ}$, which is currently at 0.710~R$_\odot$ in this model).}
%
%______________________________________________________________
 
The gravitational force exerted on an ion species ``$i$'' is $A_i\,m_H\,
g(r)$, where
$g(r)$ is the gravitational acceleration at distance $r$ from the center
of the Sun.
Radiative pressure is included in the modeling of diffusion
by changing the effective gravity felt by different
atomic species to $A_i\,m_H\, (g-g_{{\rm rad};i})$,
where $g_{{\rm rad};i}$ is the radiative acceleration.
$g_{{\rm rad};i}$ is very sensitive to the
to features of the species spectrum, i.\,e.\,to the atomic
structure of the ion,
on the ionization state of the ion and of other ions
competing for the same photons. Completely ionized elements feel
practically no radiation acceleration.
 
The expression for $g_{{\rm rad};i}$ is taken from {\it Richer et
al.} [1998]
(please refer to this paper for a detailed discussion)
as being
\begin{equation}
  g_{{\rm rad};i} = \frac{L_r}{4\pi\,r^2\,c} \frac{\kappa_R}{ X_i} \,
f_{\rm flux} \, .
\end{equation}
In this expression $X_i$ is the mass fraction of species $i$, $f_{\rm
flux}$
is the fraction of the radiative flux which will be transferred to
species $i$
(refer to Eq.1 in {\it Richer et al.}~[1998]).
The Rosseland mean opacity $\kappa_R$ is a measure of the opacity of the
plasma and determines the
total momentum of the radiation field transferred to the absorbing
plasma.
 
The inverse relationship between $g_{{\rm rad};i}$
and the species abundance has an interesting
potential consequences for isotopic fractionation.
Different isotopes of an element can have large differences in
abundance, which would favor the less abundant species.
This could potentially create large isotopic anomalies. 
The condition for such an effect is that the spectral lines of the 
minor isotope be well separated from the spectral lines of the major isotope
and that there would be no coincidence with strong lines from
other elements so that $f_{\rm flux}$ can become non negligible for the
minor isotope.
This is not the case in the Sun, not even for helium for which
the effect would be the most noticeable. The temperatures
below the convection zone where separation by radiative pressure 
could occur are large enough, $T > 2\times 10^6$~K, that
the thermal width of the spectral lines is much larger than 
the line separation for different isotopes of a given element.
Therefore, $f_{\rm flux}$ will be small and will cancel out the
effect of the $1/X_i$ factor.

It follows that in all calculations of the
evolution of isotopic ratios in the present paper the $g_{{\rm rad};i}$
will be assumed to be the same for all isotopes of a given element.
The radiation pressure will only play a role in the evolution of
elemental ratios.
 
\subsection{Mixing below the solar convection zone}\label{sec:mixing}
 
The first evidence for mixing outside of the convection zone 
came from the strikingly low photospheric
lithium abundance. Compared to its meteoritic abundance
it is depleted by a factor of 137
in the photosphere [{\it Carlsson et al.}, 1994].
Solar models without mixing cannot account for this large difference.
Modern solar models include various forms of mixing, based on
convective overshoot or rotationally-driven mechanisms, and can now
account for the observed lithium depletion while preserving
the observed beryllium abundance and $^3$He/$^4$He ratio
[See e.\,g.\,{\it Brun et al.}, 2000; {\it Richard et
al.}, 1996].
 
Moreover, based on helioseismic results, it appears that diffusion alone is
not sufficient
to explain the radial profile of the speed of sound in the Sun.
There are large discrepancies between older models and measured values, especially around the
lower boundary of the convection zone.
This can be and has been rectified by including a limited
extension of mixing below the convection zone into the stable radiative
interior 
[{\it Christensen-Dalsgaard}, 1996]. 
 
Both issues, lithium and sound speed, can be resolved simultaneously
by the same mixing mechanism [as in {\it Brun et al.}, 2000]
The mixing occurs in a narrow
region at the base of the convection zone in which there is a large
gradient in rotational velocity, this region is called the tachocline.
Although several mixing processes in that region can match the
constraints in the Sun, studies in other solar type stars [{\it Piau and
Turck-Chi\`eze}, 2002] suggest that
turbulent mixing caused by the shear due to the rotational velocity gradient
[{\it Spiegel and Zahn}, 1992] is superior to other processes, the most
often suggested being convective 
overshoot [e.g.\ {\it Bl\"ocker et al.}, 1998].
 
%%%%%%%%%%%%%%%%%%%%%%%%%%%%%%%%%%%%%%%%%%%%%%%%%%%%%%%%%%%%%%%%%%%%%%%%%%%%%%%%%%
\section{The predicted evolution of the composition of the outer convection zone}\label{sec:evol}
 
For this work, we have used the models calculated by {\it Turcotte et
al.} [1998] which feature the most detailed treatment of microscopic diffusion in the Sun.
The evolution of the abundance of the minor isotopes of elements 
heavier than carbon have been calculated explicitly.
The {\it Turcotte et al.} [1998] models have also been used by {\it Bochsler}
[2000] to investigate the effect of isotopic fractionation in the Sun
on solar wind abundances. He estimated the fractionation based on
isotopic mass differences and the published factors of depletion for the
major isotope of each element. In the current paper the full effect of
the solar evolution on the net isotopic fraction is taken into account
and an approximate prescription for the effect of mixing beyond the
convection zone is also used. The net effect on isotopic ratios is quite
similar in {\it Bochsler} [2000] and here in the absence of mixing.
While the {\it Turcotte et al.} [1998] models do not include the effect 
of mixing beneath the convection zone,
we have approximated how mixing affects the evolution of the
photospheric composition by reducing the changes in composition such as to reflect
the results of {\it Brun et al.} [2000] who include mixing in the
tachocline in addition to a standard treatment of diffusion.
 
Because of the assumptions on initial composition inherent in the
computation of solar models, the absolute values of the
abundances predicted by solar evolution models cannot be used,
only relative changes
are meaningful (with the notable exception of the helioseismic
measurements of the helium abundance).
Here we assume an initial composition as in {\it Grevesse and Noels}
[1993] calibrated to reproduce a contemporary value of $Z/X$ of 0.245. 
This value of $Z/X$ inferred from the observations
is thought to be accurate to 10\% at best.
 
\begin{figure}
     \epsfig{file=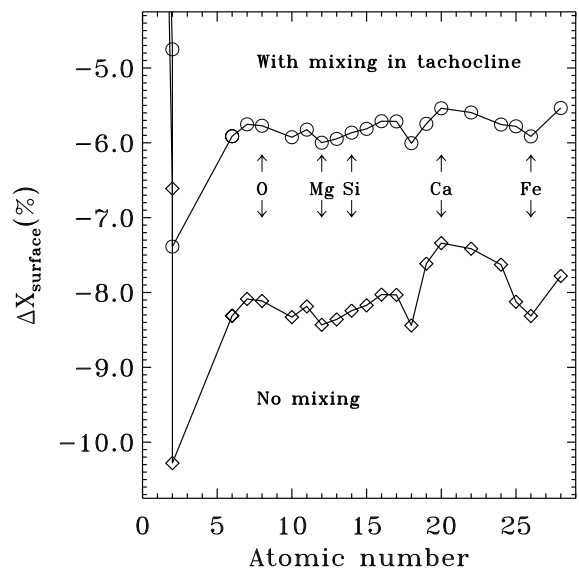,width=8cm}
     \caption{Relative change in the surface abundances predicted
                 by the detailed solar models with diffusion, the lower
                 one without mixing and the higher one with an
                 approximate
                 affect of mixing at the base of the convection zone.
              }
         \label{fig:surface}
   \end{figure}

The predicted surface composition at 4.6~Gyr is illustrated in
Fig.~\ref{fig:surface}
where the difference between the composition at 4.6~Gyr and the
original composition is plotted in function of the atomic number.
The features reflect mostly the effect of the differential radiative
forces.
These effects are relatively small however as the spread in the relative
abundance
variations spans 8.8\% for Ar to 7.6\% for Ca. The mean variation
in the detailed model is around 8.5\%. The model including mixing has a
lower relative change,
with an adopted value for all estimates done in this work
of 71.8\% as high as that of the detailed diffusion-only model.
 
%----------------------------------------------------------- surface
% {\bf Caption for Fig~3} {\em Relative change in the surface abundances predicted
% by the detailed solar models with diffusion, the lower
% one without mixing and the higher one with an approximate
% affect of mixing at the base of the convection zone.}
%
%______________________________________________________________

\begin{figure}
     \epsfig{file=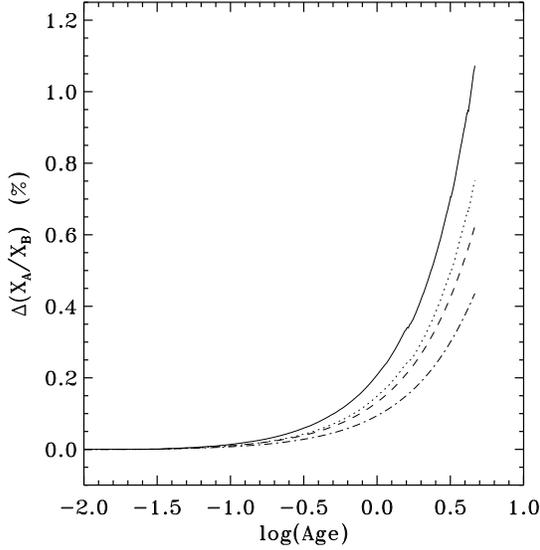,width=8cm}
      \caption{The evolution of the calcium over iron  and
               $^{12}$C/$^{13}$C abundance ratios
               in the solar convection zone over it's lifetime.
               The solid line and dotted lines show Ca/Fe with and
               without mixing below the convection zone respectively.
               The dashed and dot-dashed lines show $^{12}$C/$^{13}$C
               with and without mixing.
              }
         \label{fig:surft}
   \end{figure}

As a consequence of the evolution of the Sun itself, reflected in the decreasing
timescales shown in Figure~\ref{fig:timescale},
the rate of abundance changes has increased sharply 
as the Sun has aged.
This is illustrated in Fig.~\ref{fig:surft}
where the elemental abundance ratio of calcium to iron 
and the isotopic abundance ratio of $^{12}$C and $^{13}$C are shown 
to increase exponentially in the near past. One notices that the trends
are the same for the elemental and isotopic ratios even though
radiative forces are irrelevant in the latter case.
One consequence of this rapid recent evolution is that fossil records
of the solar wind at different epochs (lunar soils for example) may
possibly contain evidence of the variations of some abundance
ratios.

%----------------------------------------------------------- surface
% {\bf Caption for Fig~4} {\em The evolution of the calcium over iron  and
               % $^{12}$C/$^{13}$C abundance ratios
               % in the solar convection zone over it's lifetime.
               % The solid line and dotted lines show Ca/Fe with and
               % without mixing below the convection zone respectively.
               % The dashed and dot-dashed lines show $^{12}$C/$^{13}$C
               % with and without mixing.}
%
%______________________________________________________________

The predicted evolution of isotopic ratios for He, C, N, O, Ne, Mg, Si,
and Ar is presented in Table~\ref{tab:isotope}.
Recalling the discussion on diffusion processes (Section~\ref{sec:evol}), one can 
infer that
the only difference of note for isotopes in the context of diffusion
is their mass. It is assumed here that
all characteristics (i.\,e.\,ionization state, radiative forces) 
for the major isotope were shared by the minor isotopes
with the exception of mass.

As a result of the large mass difference between its isotopes, 
helium has the largest variation in its isotopic ratio. 
Although the variation of the isotopic ratio is by far the largest
variation of all ratios either isotopic or elemental, the study of helium
is compromised by the lack of accurate measurements of the isotopic ratio 
of helium at earlier times in the Sun's life, as has been touched upon
in the introduction. 
For other elements,
the isotopic ratios vary much less but not much more so than elemental
ratios as seen in Table~\ref{tab:ratios}. Not surprisingly, 
the largest effect, after helium, was found
for $^{18}$O/$^{16}$O and $^{36}$Ar/$^{40}$Ar which have similar
isotopic relative mass differences. 
As a general rule, as the mass of the element increases the effect decreases
gradually as the relative mass difference between isotopes decreases. 
Nonetheless, significant effects can be seen when comparing the more massive
isotopes of Mg and Si to their major isotopes.
 
\begin{table}
\begin{flushleft}
\caption{Values of the relative variation in isotopic abundance ratios over the 
         Sun's life ($\Delta(A/B)/(A/B)_o$) in percent. The value without mixing
         is taken from the {\it Turcotte et al.}~[1998] models and the
         values with mixing are the same results corrected by the estimated
         effect of tachocline mixing as discussed in {\it Brun et al.}~[1999].
         \label{tab:isotope}}
\medskip
\begin{tabular}{c|c|c}
\strut  $A/B$                & No Mixing (\%) & With mixing (\%)  \\
\hline
\strut $^3$He/$^4$He         & 4.10           & 2.85              \\
\strut $^{13}$C/$^{12}$C     & 0.61           & 0.43              \\
\strut $^{15}$N/$^{14}$N     & 0.44           & 0.31              \\
\strut $^{17}$O/$^{16}$O     & 0.37           & 0.26              \\
\strut $^{18}$O/$^{16}$O     & 0.91           & 0.64              \\
\strut $^{21}$Ne/$^{20}$Ne   & 0.26           & 0.19              \\
\strut $^{22}$Ne/$^{20}$Ne   & 0.71           & 0.50              \\
\strut $^{25}$Mg/$^{24}$Mg   & 0.19           & 0.13              \\
\strut $^{26}$Mg/$^{24}$Mg   & 0.57           & 0.40              \\
\strut $^{29}$Si/$^{28}$Si   & 0.13           & 0.09              \\
\strut $^{30}$Si/$^{28}$Si   & 0.45           & 0.31              \\
\strut $^{36}$Ar/$^{40}$Ar   & 1.09           & 0.76              \\
\strut $^{38}$Ar/$^{40}$Ar   & 0.64           & 0.43              \\
\end{tabular}
\end{flushleft}
\end{table}

%%%%%%%%%%%%%%%%%%%%%%%%%%%%%%%%%%%%%%%%%%%%%%%%%%%%%%%%%%%%%%%%%%%%%%%%%%%%%%%%%%%
\section{Selecting the best tracers of the solar chemical evolution}\label{sec:frac}
 
The aim of this work is to identify elemental or isotopic abundance ratios that
are especially susceptible to elemental migration effects, well determined in
meteorites, and which can be precisely measured in the solar wind. In this section
we consider possible biases and fractionation processes which need to be taken into
account in both meteoritic and solar wind studies and then motivate out
selection.

\subsection{Elemental Abundances From Meteorites}
\label{sec:el_met}
 
\begin{figure}
  \begin{center}
    \epsfig{file=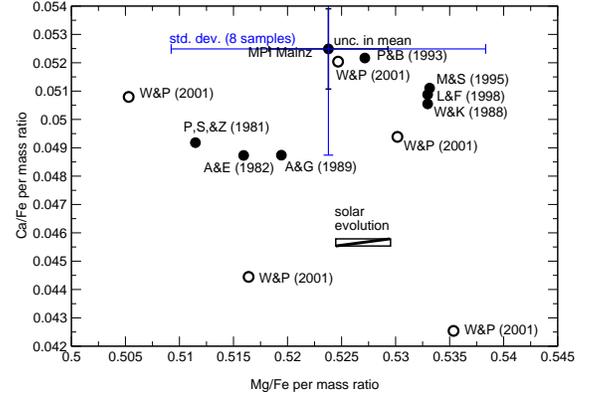,width=8cm}
    \caption{Ca/Fe vs. Mg/Fe elemental mass abundance ratios for various
             studies of CI-chondritic samples. The abbreviations
             next to the symbols refer to the authors of the
             corresponding studies: P,S,\&Z (1981), Palme, Suess, and Zeh, (1981);
             A\&E (1982), Anders and Ebihara (1982); A\&G (1982), Anders
             and Grevesse (1982); W\&K (1988),
             Wasson and Kallemeyn (1988); L\&F (1998), Lodders and Fegley, (1998); M\&S
             (1995), McDonough and Sun, (1995); P\&B (1993), Palme and Beer, (1993); MPI
             Mainz, Spettel et al., (1993). We have also plotted the standard deviation of MPI
             measurements of 8 samples as well as the uncertainty in the
             unrealistic mean. Single measurement errors are smaller
             than this uncertainty, the scatter reflects variability
             in different mineral separates of CI chondrites Orgeuil and
             Alais. The diagonal of small box shows the effect of
             solar evolution on photospheric abundances. The absolute
             position of the box is unimportant, it describes the
             relative change of the two elemental abundance ratios.
             \label{fig:Mg_Ca_Fe_CI_only}}
  \end{center}
 \end{figure}

Because meteorites can be analyzed in the laboratory, elemental
abundance measurements
performed on meteorites or terrestrial samples reach the highest
possible precision.
Measurement uncertainties are vanishingly small even for the application
considered in this work.
In Figure~\ref{fig:Mg_Ca_Fe_CI_only} we have plotted the Ca/Fe
element abundance ratio versus the Mg/Fe ratio as derived by different
authors (indicated next to the symbols and referenced in the caption).
The actual measurement
uncertainties are considerably smaller than the spread in all the data
and also smaller than the spread in the data
from the Mainz group, who publish abundance data for 8 samples. Their
reported uncertainty in the mean is about 1\% for Mg/Fe and larger for
Ca/Fe. It is important to note that these uncertainties are not
primarily due to the different authors of the studies, but rather  to
different (some subjective) choices
of samples of meteoritic matter. Moreover, there are only five known CI
chondrites, of which one has been
studied extensively (Orgeuil), one in some detail (Alais), and three are
so small that obtaining samples from
them must be considered a feat on its own. Hence the observed variations
in samples in Figure \ref{fig:Mg_Ca_Fe_CI_only}
must be considered a lower limit on the variability of these elemental
abundance ratios in different separates of
the most faithful samples of the presolar nebula.
For comparison, we have also plotted a box with a diagonal which shows
the maximum evolutionary change of
the photospheric abundance ratios
based solar evolution models without mixing. Obviously, even if solar
wind measurements were possible with
much better precision than currently possible (for instance with the
Genesis samples), the effect of solar
evolution can not be determined by considering elemental abundance
ratios. Uncertainties in meteoritic
abundances will render futile any such endeavor.
 
%-------------------------------------------------------------------- 
% {\bf Caption for Fig~5} {\em Ca/Fe vs. Mg/Fe elemental mass abundance
% ratios for various studies of CI-chondritic samples. The abbreviations
% next to the symbols refer to the authors of the
% corresponding studies: P,S,\&Z (1981), Palme, Suess, and Zeh, (1981);
% A\&E (1982), Anders and Ebihara (1982); A\&G (1982), Anders
% and Grevesse (1982); W\&K (1988),
% Wasson and Kallemeyn (1988); L\&F (1998), Lodders and Fegley, (1998); M\&S
% (1995), McDonough and Sun, (1995); P\&B (1993), Palme and Beer, (1993); MPI
% Mainz, Spettel et al., (1993); W\&P (2001), Wolf and Palme, (2001). 
% We have also plotted the standard deviation of MPI
% measurements of 8 samples as well as the uncertainty in the
% unrealistic mean. Single measurement errors are smaller
% than this uncertainty, the scatter reflects variability
% in different mineral separates of CI chondrites Orgeuil and
% Alais. The diagonal of small box shows the effect of
% solar evolution on photospheric abundances. The absolute
% position of the box is unimportant, it describes the
% relative change of the two elemental abundance ratios.}
%-------------------------------------------------------------------- 

\subsection{Isotopic Composition From Meteorites}
\label{sec:iso_met}
 
It is well known from studies such as the one of {\it Anders and Grevesse} [1989] % AG89
that there is little
variation in the abundance ratios among the non-volatile elements in
different  materials of the solar system.
This homogeneity in elemental compositions of different solar system
materials reflects the fact that
the early condensates in the inner parts of the solar nebula did not
undergo strong chemical fractionation.
Isotopic fractionation - e.\,g.\,due to incomplete condensation - was
limited to tiny effects [{\it Humayun and Clayton}, 1995b].
Consequently, it is not surprising that there is also not much variation in
the isotopic abundance of the refractory elements in the bulk of solar
system samples. 

%-----------------------------------------------------------------------
 \begin{figure}
  \begin{center}
   \epsfig{file=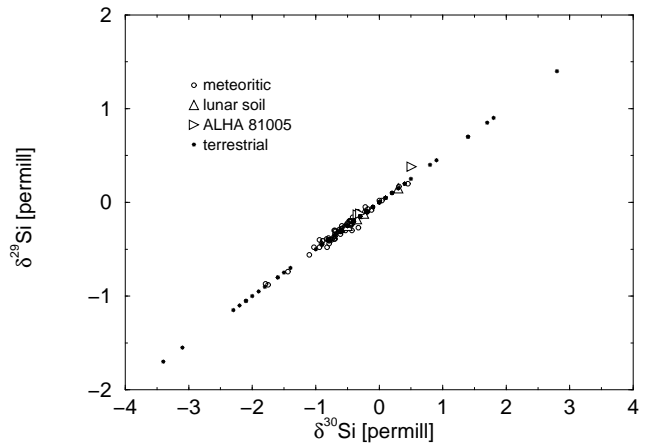,width=6cm,angle=270}
  %\vskip1.5in
  \caption{Compilation of the silicon isotopic composition of various
           sources in the solar system.
           See {\it Wimmer-Schweingruber et al.}~[1998] for sources and detailed discussion.
           \label{fig:sol_sys}}
  \end{center}
 \end{figure}
%{\bf Caption for Fig~6} {\em Compilation of the silicon isotopic
%composition of various sources in the solar system.
%See {\it Wimmer-Schweingruber et al.}~[1998] for sources and detailed discussion.}
%-----------------------------------------------------------------------

The small variation of solar system isotopic abundances in refractory
elements,
is illustrated in Figure~\ref{fig:sol_sys} for the case of Si. The conventional 
$\delta$ notation is used, $\delta^{29}$Si 
$=\,1000\,\times\,((n^{29}{\rm Si}/n^{28}{\rm Si})_{\rm sample}/(n^{29}{\rm Si}/n^{28}{\rm Si})_{\rm ref} - 1)$ 
in permill. In this 
three-isotope diagram
a variety of different solar system samples follow a straight line with
slope 1/2 which is typical for
chemical isotope fractionation. The entire variability is confined to a
narrow range of few permills per mass unit and
hence the uncertainty of the mean of the $\sim 200$ samples is much smaller, about .15 permill. 
This clearly reflects the
efficient isotopic homogenization of the solar nebula. It also lends
confidence to the assumption that there cannot be
much difference between the bulk isotopic composition of Si and other
refractory or weakly volatile
elements in the Sun and the rest of the solar system samples, including
terrestrial ones. It is for this reason that
Si and Mg are often used as standards for isotopic fractionation in the
solar wind (to within few permill per mass unit).

\subsection{Elemental Composition of the Solar Wind}
\label{sec:sol_wind_el}
 
The composition of the solar corona is not the same as that of the
photosphere and the differences
allow us to characterize various fractionation processes. There are
several properties of
an individual atom or ion that allow for fractionation. In the strong
gravitational field of the Sun,
fractionation according to mass would be extremely strong if there were
no other forces acting
on the particles. Simplistically assuming an isothermal atmosphere
between the photosphere and
the chromosphere, a barometric equation would tell us that we should not
be observing any heavy
elements in the corona or the solar wind at all. The exponential nature
of this equation
thoroughly depletes any mass but the lightest. Obviously, this is not
the case, and other forces
must be acting upon the ions and atoms already in the lowest layers of
the solar atmosphere.
Among the best known is the process that, in some still badly understood
way, separates neutral
and charged particles. Physically, it is probably based on a competition
of two time scales, the
first ionization time (FIT) and some time scale that characterizes the
removal or supply of neutral
material. Because FIT is a model dependent quantity, it is more
conventional to speak of this
important fractionation effect as the FIP (first ionization potential)
effect. The FIP is a well
determined atomic property for all important elements in the solar wind.
 
In order to describe the FIP effect, it has become customary to use a
double ratio of abundances.
One normally determines the solar wind abundance ratio of element X with
respect to oxygen, X/O,
and normalizes this observed quantity with the corresponding
photospheric value, X/O$_{\rm photo}$.
The double ratio $\Xi = $ (X/O)/(X/O$_{\rm photo}$) lies between 2 and 3
for low-FIP elements
(FIP $< 10$ V) in the slow wind and between 1 and 2 in the fast wind
[e.\,g.\,{\it von Steiger et al.}, 2000]. 
Because volatility is generally correlated with FIP this means that
refractory elements (i.\,e.\,low-FIP)
are enhanced by a factor of about two in the average solar wind when
compared with their photospheric abundance.
 
Because of the magnitude of the FIP effect, only high-speed solar wind
should be used for testing solar evolution
models. However, even in this otherwise benign type of solar wind,
elemental abundances are not necessarily
fully determined by the FIP effect. Especially wave-particle
interaction, widely believed to be the accelerating agent
in this type of solar wind, is highly dependent on the ionization state
of the ion and hence on its atomic structure.
Given current uncertainties in atomic parameters, it appears unlikely
that elemental abundances will be used in
the near future for the purpose studied in this work.

\subsection{Isotopic Composition of the Solar Wind}
\label{sec:sol_wind_iso}

The strong fractionation observed for elements in the slow solar wind is
mostly due to the FIP effect
(First Ionization Potential). Although this effect is of utmost
importance for elemental fractionation,
it is probably quite unimportant for isotopic fractionation. For
instance, the elemental fractionation
model of {\it Marsch et al.}~[1995] can be understood as a competition
between two time scales, diffusion and ionization.
The authors derive the following expression for the mass-dependent part
of
elemental fractionation $f_{ij}$ between two species $i$ and $j$,
\[f_{ij} \propto \left( \frac{A_i + 1}{A_i} \frac{A_j}{A_j +
1}\right)^{1/4}, \]
where $A_{i,j}$ is the atomic mass number of species $i$ and $j$. For
the example of magnesium this
expression evaluates to a depletion of the heavy isotopes of less than
half a permill per mass unit.
In other words, FIP is unimportant for isotopic fractionation according
to the {\it Marsch et al.}~[1995] model. Note however,
that the model of {\it Schwadron et al.}~[1999], which invokes resonant
ion-cyclotron heating of
weakly ionized species to explain the FIP effect, could produce
significantly stronger isotope effects. This is
concluded from analogy with the severe isotope fractionation observed in
energetic particles
related with impulsive flare events.
 
Inefficient Coulomb drag is most probably the most efficient isotopic
fractionation process.
Under the assumption of an isothermal solar atmosphere and neglecting
wave acceleration, one may derive the following
expression for the maximum fractionation  between the two isotopes $k$ and $l$ of an
element due to inefficient Coulomb drag
[{\it Bodmer}, 1996; {\it Bodmer and Bochsler}, 1998; {\it Bodmer and Bochsler}, 2000]
\[ f_{kl} \propto \frac{2 A_k - Q - 1}{2 A_l - Q - 1} \sqrt{\frac{A_k
+1}{A_l + 1}}. \]
This evaluates to more than 5\% per mass unit for the example of silicon. Thus,
for isotopic fractionation, inefficient Coulomb
drag can be two orders of magnitude larger than the FIP effect.
The fractionation of isotopes has been tested for in the solar wind.
Combining measurements of Ne, Mg, and Si,
{\it Kallenbach et al.}~[1998b] found a depletion of the heavier isotopes in
the slow solar wind.
However, the isotopic composition of the solar
wind is consistent with meteoritic values in high-speed streams
[{\it Bochsler et al.}, 1997, {\it Kallenbach et al.}, 1997c].
This confirms theoretical
expectations that isotopic fractionation should be very weak in
high-speed streams
[{\it Bodmer and Bochsler}, 1998; {\it Bodmer and Bochsler}, 2000].
Since the proton flux is high in the
high-speed solar wind,  Coulomb friction is probably unimportant in
these streams, consistent with UVCS measurements of
high outflow velocities of heavy ions in coronal
holes~[e.\,g.\,{\it Kohl et al.}, 1997]
which are the source regions of the high-speed streams
[{\it Krieger et al.}, 1973].
Since these observations show oxygen out-flowing at higher speeds than
protons [{\it Cranmer et al.}, 1999],
it appears that Coulomb drag
plays no role in carrying heavy species in the high-speed solar wind. If
indeed wave-particle interaction succeeds to impart
the momentum needed to move these species into the coronal hole
associated solar wind, as is generally believed,
isotope effects are expected to be weak. This is also borne out by the
observed constancy of the $^3$He/$^4$He isotopic
abundance ratio in high-speed streams
[{\it Bodmer and Bochsler}, 1998; {\it Gloeckler and Geiss}, 1998].
It is thus to be expected that the isotopic composition
of the fast solar wind most accurately resembles that of the reservoir
that it is being fed from.
Hence it is measurements in high-speed streams which should be used to
limit the amount of
gravitational settling of the heavy isotopes toward
the solar core during solar evolution.

\subsection{Selection}
\label{sec:selec}
 
In view of the variations in meteoritic elemental abundance ratios and
in the expected magnitude of
solar wind fractionation processes, it appears unrealistic to use
elemental abundance ratios as a test
for solar evolution  models. Nevertheless, for completeness sake, we
give a list in Table~\ref{tab:ratios}
of those elemental abundance
ratios for low-FIP ($< 10$ V) elements for which the effect of solar
evolution exceeds 2.5 permill. Of these
Ca/Mg is the top candidate from the evolution point of view, but, as
already mentioned, is not determined
with sufficient accuracy in CI chondrites. If some clear-cut criterion
for sample selection could be found
that would narrow down the sample variability, this could be the best
compositional test of solar evolution models.
 
On the other hand, the small variations in the isotopic composition of
refractory
elements (e.\,g.\,Mg, Si,\ldots) obvious from the already discussed Figure~\ref{fig:sol_sys}
should allow such a test. Current in-situ
mass spectrometers are not capable of measuring the isotopic composition
of the solar wind to the precision
necessary for this study. Linear time-of-flight mass spectrometers do
not resolve isotopes of refractory elements
and isochronous mass spectrometers have systematic uncertainties which
are about two orders of magnitude too large
to detect the effect of solar evolution. However, the recently launched
Genesis mission will collect and return to
Earth samples of the solar wind in various solar wind regimes including
the fast solar wind. These samples will
be available to laboratory analysis. The Genesis mission has the aim of
measuring the isotopic composition of oxygen
to a relative precision of $10^{-4}$. The abundance of the rare isotopes
of oxygen are low ($^{17}$O/$^{16}$O $\sim 1/2625$,
$^{18}$O/$^{16}$O $\sim 1/499$) in solar system samples 
[{\it Anders and Grevesse}, 1989; {\it Clayton}, 1993] and in the solar
wind [{\it Collier et al.}, 1998; {\it Wimmer-Schweingruber et al.}, 2001].
Because of the large abundance of the rare isotopes of Mg and Si
isotopes, this measurement should be possible with these
samples. 
This measurement precision for O isotopes translates to a similarly high
precision for Mg and Si isotopes.
Because the heavy and less abundant isotopes of Mg and Si are not nearly
as rare as those of O, Genesis should
return high-precision and statistically significant samples of the
isotopic composition of Mg and Si in the solar wind.
 
Mg and Si have other positive aspects to them that make them attractive
for this study. The are the lightest refractive
elements with three stable isotopes, moreover, their rare isotopes are
still fairly abundant. In addition, as can be seen
in Figure~\ref{fig:surface}, they lie in a local minimum in the
predicted photospheric abundance $\Delta$X$_{\rm surface}$,
i.\,e.\,a local maximum in the change. Because the isotopic effect is
well approximated by $\Delta$X$_{\rm surface}$ times
the relative mass difference of the isotopes, the isotopes of these two
elements offer the best chance of measuring subtle
changes in their composition due to solar evolution.
 
\begin{table}
\begin{flushleft}
\caption{Some of the highest photospheric abundance ratios as
predicted by solar models for elements of FIP  $<$ 10eV.
The absolute values of the difference in mass and FIP
as well as of the relative change in abundance ratios
at the solar age. The numeral on the leftmost
column indicates the position in the list of all
relative variations of the ratios in inverse
numerical order.\label{tab:ratios}}
\begin{tabular}{cccccc}
order   & A/B   & $\Delta$(mass)& $\Delta$(fip) & \multispan 2 \hfill
$\Delta(X_A/X_B)$ (\%) \hfill  \\
        &       & (amu)         & (eV)          & No mixing & Mixing  \\
\hline
2       & Ca/Mg & 15.8          & 1.53          & 1.19      & 0.84
\\
4       & Ca/Al & 13.3          & 0.13          & 1.11      & 0.78
\\
5       & Ti/Mg & 23.6          & 0.83          & 1.11      & 0.78
\\
7       & Fe/Ca & 15.8          & 1.76          & 1.05      & 0.74
\\
9       & Ti/Al & 21.1          & 0.83          & 1.03      & 0.72
\\
11      & Ca/Si & 12.0          & 2.04          & 0.98      & 0.69
\\
12      & Fe/Ti &  8.0          & 1.05          & 0.97      & 0.68
\\
14      & Ca/Na & 17.1          & 0.97          & 0.92      & 0.65
\\
16      & Ti/Si & 19.8          & 1.33          & 0.90      & 0.63
\\
18      & K/Mg  & 14.8          & 3.31          & 0.89      & 0.62
\\
20      & Cr/Mg & 27.8          & 0.88          & 0.88      & 0.61
\\
21      & Mn/Ca & 14.9          & 1.32          & 0.85      & 0.59
\\
25      & K/Al  & 12.3          & 1.65          & 0.81      & 0.57
\\
29      & Mn/Ti &  7.1          & 0.62          & 0.77      & 0.54
\\
31      & Fe/K  & 16.7          & 3.53          & 0.76      & 0.53
\\
36      & Fe/Cr &  3.9          & 1.10          & 0.74      & 0.52
\\
51      & Ni/Fe &  2.8          & 0.24          & 0.58      & 0.41
\\
\hline
\end{tabular}
\end{flushleft}
\end{table}
 
%%%%%%%%%%%%%%%%%%%%%%%%%%%%%%%%%%%%%%%%%%%%%%%%%%%%%%%%%%%%%%%%%%%%%%%%%%%%%%%%%%%%%%%%
\section{Discussion}
\label{sec:disc}

Today standard solar models include effects of elemental segregation. In
this picture,
all elements heavier than hydrogen are gradually depleted in the solar
photosphere over the
course of solar evolution. Their current photospheric abundances are
about 8\% lower than
in the bulk Sun or in the presolar nebula. This depletion is primarily
due to differences in the
radiative and gravitational forces acting on the various elements. Such
models compare favorably with
sound speed profiles of the Sun inferred from helioseismology.
Nevertheless, the agreement
is far from satisfactory, differences in inferred and model sound speeds
are still very large compared
to the modeled uncertainties on their magnitudes.
 
In this paper we have investigated a more direct
approach to obtain limits on the amount of element segregation
throughout solar history.
The smallness of the effect makes it very difficult to detect. There are
two possibilities to check
such a secular change in photospheric abundances. One can measure
implanted solar wind in lunar soils of
various antiquity, or one can compare contemporary photospheric or solar
wind abundances with bulk solar
abundances inferred from (predominantly) meteoritic data. The first test
has proven to remain ambiguous
[{\it Heber}, 2002].
Uncertainties in photospheric abundance
determinations are on the order of or exceed 25\% - 30\%. These
uncertainties  are mainly due to badly known
atomic properties [{\it Del Zanna et al.}, 2001], but also due to the badly
understood physics of the photosphere and lower chromosphere
[{\it Holweger}, 2001; {\it Del Zanna et al.}, 2001]. Hence it is not to be expected
that photospheric abundances will be measured in the near
future
with sufficient accuracy to measure the effect of element migration in
the near future.
Therefore, we have investigated the possibility of detecting or at least
putting limits on this effect by
using in-situ solar wind data.
 
The use of in-situ solar wind data immediately restricts the set of
elements than may be considered. Because of the magnitude
of the insufficiently understood FIP effect, only elements with low FIP
(FIP $< 10$ V) should be considered for comparison.
However, in this case, no comparison can be made with hydrogen, implying
that the magnitude of the effect of element settling
is decreased considerably. It amounts to only typically 0.5\% for most
element combinations. We give a list of the
best candidate pairs in Table~\ref{tab:ratios}. The smallness now results
in an additional difficulty, namely that
the normalizing
meteoritic abundances (or element ratios) are not known with sufficient
accuracy.
 
We have shown that only isotopes of refractory elements offer any chance
to measure the magnitude of the effect of
elemental segregation
in the solar convection zone. With Genesis it
should be possible to determine the Mg and Si isotopes with an accuracy
of about 1 permill, good enough to detect this effect
because in this case, meteoritic data are of comparable or better
quality.
 
Alternatively, high-precision measurements of solar wind noble gas 
isotopic abundances in lunar soils still has the potential to unravel
solar chemical evolution if isotopic and elemental fractionation in the
implantation process can be reliably modeled
and if it can be understood in sufficient detail throughout the complex
history of lunar soils and other archives for
the solar wind.
 
The alteration of the photospheric (and hence solar wind) isotope
composition is especially important for oxygen which will
be measured at a very high precision ($\sim$ 0.01\%) with Genesis. Here,
accurate measurements of
isotope ratios of refractory elements such as Mg and Si will be
important to ``calibrate'' the influence of solar evolution on
more volatile and chemically reactive elements such as oxygen. Inclusion
of element migration throughout solar history will
be crucial to the interpretation of the measured values of
$^{16}$O/$^{18}$O and $^{16}$O/$^{17}$O. Because solar
isotopic evolution will tend to follow the traditional chemical fractionation line slope of
$\sim 1/2$ while alternative models for O isotopes in the solar system involve
non-mass-fractionating processes, a precise knowledge of the solar
effects is of utmost importance for the understanding of the history of
the solar system.

While it is doubtful that the required understanding of all fractionation
processes in the solar wind exists at this time for an accurate analysis of 
isotopic ratios in relation to physical processes in the Sun but we
believe that the necessary knowledge is attainable.

\begin{acknowledgments}
This work was performed in part under the auspices of the U.S.
Department of Energy, National Nuclear Security Administration by the University of California,
Lawrence Livermore National Laboratory under contract No. W-7405-Eng-48
and by the Swiss National Science Foundation.
\end{acknowledgments}

\end{article}
\end{document}